\documentclass[prl,showpacs,twocolumn]{revtex4}

\usepackage{graphicx}

\begin{document}
\title{Quantum-noise-limited Angular Momentum Measurement \\ for a Micron-sized Dielectric Object}

\author{Koji Usami$^{1,2}$} \email{usami.k.ab@m.titech.ac.jp}

\affiliation{$^1$PRESTO, Japan Science and Technology Agency, 3-5, Sanbancho, Chiyodaku, Tokyo 332-0012, Japan \\
$^2$Department of Physics, Tokyo Institute of Technology, 2-12-1 O-okayama, Meguro-ku, Tokyo 152-8550, Japan}
\date{\today }

\begin{abstract}
An approach is described for observing quantum features of micron-sized spinning objects. Specifically, we consider a birefringent (uniaxial positive) dielectric object in the shape of an oblate (i.e., frisbee-like) symmetric top. It can be trapped in the air, its extraordinary axis can be aligned, and its angular momentum along the extraordinary axis can be stabilized, all optically. We show that the angular momentum quantum noise of the object perpendicular to the gigantic angular momentum along the extraordinary axis can be measured as a linear birefringent phase shift of a probe laser in an analogous fashion to the spin quantum nondemolition (QND) measurement in atomic physics.  
\end{abstract}

\pacs{03.65.Ta,42.50.Dv,42.50.Wk} 

\maketitle

Whether all objects, including macroscopic ones, conform to quantum mechanics or not~\cite{Leggett1980ptp_s} is still at issue. Thanks to the technological advance, a bunch of groups started to explore the issue experimentally~\cite{Leggett2002jpc}. The recent spectacular achievements in this direction include the reduction of the phonon number in a nanomechanical resonator mode~\cite{NBLACBS2007n} and the detection of the plasmon excitation number in a superconducting circuit~\cite{SHSWGBFMJDGS2007n}, just to name a few. These objects are macroscopic in a sense that they (e.g., the nanomechanical resonator~\cite{NBLACBS2007n} and the superconducting circuit~\cite{SHSWGBFMJDGS2007n}) are constructed out of a large number of particles like electrons and nucleons, yet their motions can be described by the collective coordinates which may eventually behave quantum mechanically~\cite{YD1984a}. To see the quantum behavior~\cite{ABS2002,MSPB2003}, these collective excitations are to be strongly coupled to the (artificial) two-level system (e.g., a Cooper pair box~\cite{NPT1999n}, or a single photon in a interferometer) and the latter essentially acts as a quantum counting device~\cite{CTDSZ1980rmp,BVT1980s} for the former.

Here, we present an alternate approach for observing quantum features of apparently classical objects. Our scheme is inspired by the spin quantum nondemolition (QND) measurement in atomic physics~\cite{KBM1998epl,KMJYEB1999a,THTTIY1999a,KMB2000,JKP2001n,JSCFP2004n,GSM2004s}, and follows the homodyne (quadrature measurement) paradigm in quantum optics~\cite{UTK2006a} instead of quantum counting (number measurement). 

\begin{figure}
\includegraphics[width=\linewidth]{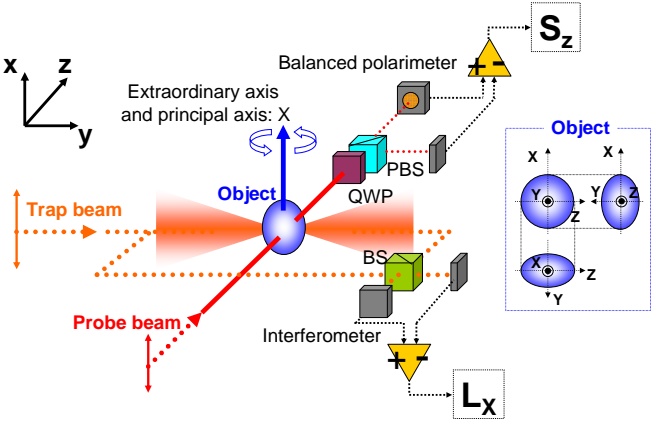}
\caption{An object is a birefringent dielectric top. Dipole force with a focused laser beam is used to trap the object in the air. Extraordinary axis, X, can be aligned along x axis by the x-polarized trap laser. Surviving angular momentum along the extraordinary axis, X, can be stabilized by monitoring the angular velocity with an interferometer and additional lasers (not shown) to add or subtract the required angular momenta. The angular momentum quantum noise perpendicular to the gigantic angular momentum along extraordinary axis, X, can be measured by a probe laser and a balanced polarimeter with a quarter wave plate (QWP) and a polarization beam splitter (PBS).}
\label{fig:schematic}
\end{figure}

Specifically, we consider a micron-sized birefringent (uniaxial positive) dielectric object in the shape of an oblate (frisbee-like) symmetric top. A schematic of the proposed experimental setup as well as the shape of the object are shown in Fig.~\ref{fig:schematic}. Hereafter, a caret ( $\hat{}$ ) will be used above a quantity to denote an operator and to be distinguished from a $c$-numbered quantity. The lower-case $xyz$ denote the axes in the space-fixed frame, whereas the capital $XYZ$ denote those in the object's body-fixed frame. The dielectric object can be trapped in the air by the dipole force with a focused laser beam~\cite{ADBC1986op,OKS1997op}, its extraordinary axis, $X$ (which is also one of the principal axes of the oblate symmetric top as shown in Fig.~\ref{fig:schematic}), can be aligned along $x$ axis by the $x$-polarized trap laser because there is the torque arisen from the difference of the refraction indices for the object, i.e., $n_{e}$ for the polarization along $X$ axis (the extraordinary axis) and $n_{o}$ for that along $Y$ axis (the ordinary axis)~\cite{FNHR1998n,BNHR2003a,LW2004}. Because the object is assumed to be an uniaxial positive ($n_{e}>n_{o}$), the expectation values of the angular momenta of the object, $\langle \hat{L}_{Y} \rangle$ and $\langle \hat{L}_{Z} \rangle$, can be both made equal to zero due to the torque~\cite{LW2004}. Besides, the surviving angular momentum, $\langle \hat{L}_{X} \rangle$, along the extraordinary axis $X$ can be estimated from the \textit{a priori}-known principal moment of inertia, $I_{X}$, and $X$ component of the angular velocity, $\hat{\omega}_{X}$, which can, as shown in Fig.~\ref{fig:schematic}, be monitored by an interferometer owing to its shape perpendicular to the extraordinary axis $X$. $\langle \hat{L}_{X} \rangle$ may be then stabilized using auxiliary lasers (not shown in Fig.~\ref{fig:schematic}) to add or subtract the required angular momenta according to the acquired angular velocity, $\hat{\omega}_{X}$. Essentially, these procedures reduce the effective entropy of the concerned degree of freedom, that is, the rotational motion of the object. The micron-sized object can thus be prepared in the state with $\langle \hat{L}_{Y} \rangle=\langle \hat{L}_{Z} \rangle=0$ yet $\langle \hat{L}_{X} \rangle \gg 1$ and $\langle (\Delta \hat{L}_{X})^{2} \rangle \equiv \langle \hat{L}_{X}^{2} \rangle - \langle \hat{L}_{X} \rangle^{2} \ll \langle \hat{L}_{X} \rangle^{2}$, which corresponds to the coherent spin state (CSS) in atomic spin systems~\cite{KBM1998epl,KMJYEB1999a,THTTIY1999a,KMB2000,JKP2001n,JSCFP2004n,GSM2004s}. For example, a oblate symmetric top made of quartz with a mass density of 2.65~g$\,$cm$^{-3}$, lengths of two orthogonal semi-major axes ($X$ and $Z$ axes) being 2~$\mu$m, and that of a semi-minor axis ($Y$ axis) being 1~$\mu$m, has a total mass, $\mu$, of about $4.44\times10^{-11}$~g and a moment of inertia, $I_{X}$, equal to $4.44\times10^{-26}$~kg$\,$m$^{2}$. Suppose that the angular velocity $\hat{\omega}_{X}/2\pi$ is stabilized to be 1~Hz, the resultant angular momentum, $\langle \hat{L}_{X} \rangle$, is as large as $2.6\times10^{9}$ in units of $\hbar$. 

Our purpose here is to show that the angular momentum quantum noise for $\hat{L}_{Y}$ or $\hat{L}_{Z}$ of the object, which is associated with the gigantic angular momentum $\langle \hat{L}_{X} \rangle$, is measurable as a linear birefringent phase shift of a probe laser in an analogous fashion to the spin QND measurement. Although there may be many technical obstacles to realize the proposed experiment, we discuss the interaction between the spinning object and the probe photons under the ideal situation. The practical issues will be discussed later.

We begin by analyzing the propagation of the probe laser through the birefringent object with Jones calculus~\cite{Fowles} to find the phenomenological evolution operator. Here, we assume that the trap laser and the auxiliary lasers for preparing the object in the CSS-like state (i.e., $\langle \hat{L}_{Y} \rangle=\langle \hat{L}_{Y} \rangle=0$ yet $\langle \hat{L}_{X} \rangle \gg 1$ and $\langle (\Delta \hat{L}_{X})^{2} \rangle \ll \langle \hat{L}_{X} \rangle^{2}$) are switched off and thus the object starts to free-fall when the probe laser begins to interact with the object. In the Schr\"{o}dinger picture, the initial state of the probe photons can be represented by
\begin{eqnarray}
&\!& |\Phi_{i}\rangle = |\beta \rangle_{x}\otimes|\gamma \rangle_{y} \nonumber \\
&=& e^{-\frac{1}{2}(|\beta|^{2}+|\gamma|^{2})}e^{(|\beta|^{2}+|\gamma|^{2})(\alpha_{x}(0)\hat{a}_{x}^{\dagger}+ \alpha_{y}(0)\hat{a}_{y}^{\dagger})}|0\rangle,
 \label{eq:initial}
\end{eqnarray}
where $|\beta \rangle_{x}$ and $|\gamma \rangle_{y}$ are the coherent states with the polarization along $x$ and $y$ axes and $\hat{a}_{x}$ and $\hat{a}_{y}$ are the creation operators for these two modes, respectively. The coefficients $\alpha_{x}(0) \equiv \frac{\beta}{|\beta|^{2}+|\gamma|^{2}}$ and $\alpha_{y}(0) \equiv \frac{\gamma}{|\beta|^{2}+|\gamma|^{2}}$ evolve into 
\begin{equation}
\left(
    \begin{array}{c}
    \alpha_{x} (l)\\
    \alpha_{y} (l)
    \end{array}
\right)
=
\left(
    \begin{array}{cc}
    e^{ik_{e}l} & 0 \\
    0 & e^{ik_{o}l}
    \end{array}
\right)
\left(
    \begin{array}{c}
    \alpha_{x} (0)\\
    \alpha_{y} (0)
    \end{array}
\right)
\equiv
\mathcal{B}\left(
    \begin{array}{c}
    \alpha_{x} (0)\\
    \alpha_{y} (0)
    \end{array}
\right), \label{eq:Jones1}
\end{equation}
after propagating through the object with a length of $l$ along $z$ axis. $k_{e}=n_{e}\omega/c$ and $k_{o}=n_{o}\omega/c$ are the angular wavenumbers for the extraordinary and the ordinary polarization components of the probe laser. The $2\times2$ matrix $\mathcal{B}$ in Eq.~(\ref{eq:Jones1}) is called the Jones matrix~\cite{Fowles}. Here, we assume that the body-fixed $X$ axis is exactly aligned parallel to the space-fixed $x$ axis. Note that the duration of the probe pulse is supposed to be far shorter than the inverse of the angular velocity $\hat{\omega}_{X}/2\pi$ of the object, and thus the length of the object, $l$, does not change so much during the propagation of the probe laser. The probe pulses can be set to propagate through the object along the body-fixed $Z$ axis. When the body-fixed $X$ axis of the object is slightly misaligned to the space-fixed $x$ axis by $\phi$ about $z$ axis, the Jones matrix $\mathcal{B}$ in Eq.~(\ref{eq:Jones1}) is modified as
\begin{equation}
\mathcal{B'}
\equiv
\left(
    \begin{array}{cc}
    \cos \phi & \sin \phi \\
    -\sin \phi & \cos \phi
    \end{array}
\right)
\mathcal{B}
\left(
    \begin{array}{cc}
    \cos \phi & -\sin \phi \\
    \sin \phi & \cos \phi
    \end{array}
\right). \label{eq:Jones2}
\end{equation}
Changing the basis of the Jones matrix, $\mathcal{B'}$, in Eq.~(\ref{eq:Jones2}) from the linear polarizations, $\hat{a}_{x}$ and $\hat{a}_{x}$, to the circular polarizations, $\hat{a}_{+}= \frac{1}{\sqrt{2}}(\hat{a}_{x}+i\hat{a}_{y})$ and $\hat{a}_{-}= \frac{1}{\sqrt{2}}(\hat{a}_{x}-i\hat{a}_{y})$, we have
\begin{equation}
\mathcal{B''}
\equiv
\exp \left( i \theta_{0} \hat{S}_{0} + i \theta \left( \cos 2\phi \hat{S}_{x} +  \sin 2\phi \hat{S}_{y} \right) \right), \label{eq:Jones3}
\end{equation}
where $\theta_{0}=\frac{k_{e}+k_{o}}{2}l$ and $\theta=\frac{k_{e}-k_{o}}{2}l$, and 
\begin{eqnarray}
\hat{S}_{0}&=& \left(
    \begin{array}{cc}
    1 & 0 \\
    0 & 1
    \end{array}
\right)= (\hat{a}_{+}^{\dagger}\hat{a}_{+}+\hat{a}_{-}^{\dagger}\hat{a}_{-}) \nonumber \\
\hat{S}_{x}&=& \left(
    \begin{array}{cc}
    0 & 1 \\
    1 & 0
    \end{array}
\right)=(\hat{a}_{+}^{\dagger}\hat{a}_{-}+\hat{a}_{-}^{\dagger}\hat{a}_{+}) \nonumber \\
\hat{S}_{y}&=& \left(
    \begin{array}{cc}
    0 & -i \\
    i & 0
    \end{array}
\right)=-i(\hat{a}_{+}^{\dagger}\hat{a}_{-}-\hat{a}_{-}^{\dagger}\hat{a}_{+}) \nonumber \\
\hat{S}_{z}&=& \left(
    \begin{array}{cc}
    1 & 0 \\
    0 & -1
    \end{array}
\right)=(\hat{a}_{+}^{\dagger}\hat{a}_{+}-\hat{a}_{-}^{\dagger}\hat{a}_{-}) \label{eq:S}
\end{eqnarray}
are the quantized Stokes parameters~\cite{Leonhardt}. Since the Jones matrix, $\mathcal{B''}$, in Eq.~(\ref{eq:Jones3}) represents the spatial translation of $\alpha_{+}(0) \equiv \frac{1}{\sqrt{2}}(\alpha_{x}(0)+\alpha_{y}(0))$ and $\alpha_{-}(0) \equiv \frac{1}{\sqrt{2}}(\alpha_{x}(0)-\alpha_{y}(0))$ along $z$ axis by $l$, the momentum operator $\hat{G}$~\cite{HSB1990a}, which is the generator for the spatial translation, is deduced by the relation, $\mathcal{B''}=\exp \left(i\frac{\hat{G}}{\hbar}l\right)$, namely, 
\begin{eqnarray}
\hat{G} &=& \hbar \left(\frac{k_{e}+k_{o}}{2} \hat{S}_{0} + \frac{k_{e}-k_{o}}{2} \left( \hat{S}_{x} +  2\phi \hat{S}_{y} \right) \right) \nonumber \\
&=& \left(\hbar k_{e} \hat{a}_{x}^{\dagger}\hat{a}_{x} + \hbar k_{o} \hat{a}_{y}^{\dagger}\hat{a}_{y}\right) + \phi \left(\hbar  k_{e}- \hbar k_{o} \right) \hat{S}_{y},  \label{eq:momentum}
\end{eqnarray}
where we assume $\phi\ll 1$. In this form the physical meanings of the terms are made clear; the first term represents the probe photon momenta while the second term denotes the momentum exchange between the probe photons and the object. Here, the object's contribution appears as the angle $\phi$ between the body-fixed $X$ axis and the space-fixed $x$ axis. 

We now bring the quantum feature of the spinning object to the angle $\phi$ in Eq.~(\ref{eq:momentum}). The angle $\phi$ can be considered as a quantum-mechanical operator, which is called the angle operator $\hat{\phi}$~\cite{BP1990a}, with a commutation relation, $\left[ \hat{\phi},\hat{L}_{z} \right]=i$. Here $\hat{L}_{z}=-i\frac{\partial}{\partial \phi}$ is the space-fixed $z$ component of the angular momentum. $\hat{\phi}$ and $\hat{L}_{z}$ are the same as a position operator $\hat{x}$ and a momentum operator $\hat{p}$. We have, on the other hand, another commutation relation, $\left[ \frac{\hat{L}_{y}}{\langle \hat{L}_{x} \rangle}, \hat{L}_{z} \right]=i$, as $\langle \hat{L}_{x} \rangle \gg 1$. The angle operator $\hat{\phi}$ can then be identified with $\hat{L}_{y}/\langle \hat{L}_{x} \rangle$. The momentum operator, Eq.~(\ref{eq:momentum}), is thus given by
\begin{equation}
\hat{G} = \hbar \left(\frac{k_{e}-k_{o}}{2} \left( \hat{S}_{x} +  \frac{2}{\langle \hat{L}_{x} \rangle} \hat{L}_{y}\hat{S}_{y} \right) \right) ,  \label{eq:momentum2}
\end{equation}
where the $\hat{S}_{0}$ term in Eq.~(\ref{eq:momentum}) is dropped because it affects trivially in the probe photons' evolution. Note that since the angular momenta, $\hat{L}_{x}$, $\hat{L}_{y}$, and $\hat{L}_{z}$, in the space-fixed frame are equal to $-\hat{L}_{X}$, $-\hat{L}_{Y}$, and $-\hat{L}_{Z}$, in the body-fixed frame, respectively~\cite{Zare}, we have $\hat{\phi}=\hat{L}_{y}/\langle \hat{L}_{x} \rangle=\hat{L}_{Y}/\langle \hat{L}_{X} \rangle$. 

The Hamiltonian for the rotational motion of the oblate symmetric top shown in Fig.~\ref{fig:schematic} is~\cite{Zare}
\begin{equation}
\hat{H}_{o}
= \hbar^{2} \left( \frac{ \hat{L}_{X}^{2}}{2I_{X}}+\frac{\hat{L}_{Y}^{2}}{2I_{Y}}+\frac{\hat{L}_{Z}^{2}}{2I_{Z}} \right)
= \frac{\hbar^{2}}{2I} \hat{L}_{0}^{2} +(\frac{\hbar^{2}}{2I_{Y}}-\frac{\hbar^{2}}{2I}) \hat{L}_{Y}^{2}, \label{eq:HamiltonianOST}
\end{equation}
where $I_{X}=I_{Z}\equiv I$ and $I_{Y}$ are the principal moments of inertia and $\hat{L}_{0}=(\hat{L}_{X}^{2}+\hat{L}_{Y}^{2}+\hat{L}_{Z}^{2})^{1/2}$ is the total angular momentum. Since the momentum operator, $\hat{G}$, in Eq.~(\ref{eq:momentum2}) and the Hamiltonian, $\hat{H}_{o}$, in Eq.~(\ref{eq:HamiltonianOST}) mutually commute, the Jones matrix $\mathcal{B''}$ in Eq.~(\ref{eq:Jones3}) can be modified to embrace these two contributions and promoted to the phenomenological evolution operator $\hat{B}$ for the total system,
\begin{equation}
\hat{B}
=\exp \left(i\frac{\hat{G}}{\hbar}l - i\frac{\hat{H}_{o}}{\hbar}\tau \right) \nonumber \\
=\exp \left( i \theta \hat{S}_{x} + i \theta' \hat{S}_{y} \hat{L}_{y}-i \chi \hat{L}_{y}^{2} \right), \label{eq:evolutionOp}
\end{equation}
where $\theta=\frac{k_{e}-k_{o}}{2}l$, $\theta'=\frac{2 \theta}{\langle \hat{L}_{x} \rangle}=\frac{k_{e}-k_{o}}{\langle \hat{L}_{x} \rangle}l$, $\chi=\hbar(\frac{1}{2I_{Y}}-\frac{1}{2I})\tau$, and $\tau$ is the period from the beginning of the object's free-fall to the end of the probe-object interaction. Here, we drop the $\hat{L}_{0}$ term in Eq.~(\ref{eq:HamiltonianOST}) because it affects trivially in the object's evolution, and we use the fact, $\hat{L}_{Y}^{2}=\hat{L}_{y}^{2}$. The form of the evolution operator $\hat{B}$ in Eq.~(\ref{eq:evolutionOp}) indicates two interesting features. First, the angular momentum, $\hat{L}_{y}=-\hat{L}_{Y}$, is the QND observable~\cite{CTDSZ1980rmp,BVT1980s} since $\left[\hat{L}_{y}, \hat{B} \right]=0$ when we assume that $\hat{L}_{x}$ is the $c$-numbered quantity as $\langle \hat{L}_{x} \rangle \gg 1$. Second, the $\hat{L}_{y}^{2}=\hat{L}_{Y}^{2}$ term in Eq.~(\ref{eq:evolutionOp}) gives rise to the so-called one-axis twisting~\cite{KU1993a}, by which the uncertainties of the angular momenta, $\hat{L}_{Y}$ and $\hat{L}_{Z}$, are redistributed and a certain angular momentum component in $Y-Z$ plane is spontaneously squeezed. 

Now, we show that the angular momentum quantum noise for $\hat{L}_{Y}$, which is associated with the gigantic angular momentum $\langle \hat{L}_{X} \rangle$, can be observed by measuring the Stokes parameter $\hat{S}_{z}$ after the evolution, i.e., the linear birefringent phase shift of the probe photons. In the Heisenberg picture, the quantized Stokes parameters, Eq.~(\ref{eq:S}), evolve from $\hat{S}_{i}^{(I)}$ into $\hat{S}_{i}^{(O)} = \hat{B^{\dagger}} \hat{S}_{i}^{(I)}\hat{B}$ ($i=x,y$ and $z$). Approximately, we have
\begin{equation}
\!\!\!
\left(
    \begin{array}{c}
    \hat{S}_{x}^{(O)} \\
    \hat{S}_{y}^{(O)} \\
    \hat{S}_{z}^{(O)} 
    \end{array}
\right)
\sim 
\left(
    \begin{array}{ccc}
    \theta^{2}+1 & 2\theta \theta' \hat{L}_{y} & -2\theta' \hat{L}_{y} \\
    2\theta \theta' \hat{L}_{y} & -\theta^{2}+1 & 2\theta \\
    2\theta' \hat{L}_{y} & -2\theta & 1
    \end{array}
\right)
\!\!\!
\left(
    \begin{array}{c}
    \hat{S}_{x}^{(I)} \\
    \hat{S}_{y}^{(I)} \\
    \hat{S}_{z}^{(I)} 
    \end{array}
\right), \label{eq:photonE}
\end{equation}
as both $\theta$ and $\theta'$ are small. Assuming that the initial state of the probe photons, Eq.~(\ref{eq:initial}), is $|\Phi_{i}\rangle=|\beta \rangle_{x}$ and thus $\langle \Phi_{i}| \hat{a}_{x}^{\dagger}\hat{a}_{x} |\Phi_{i} \rangle =|\beta|^{2}\equiv n$ with $n \gg1$, the annihilation operator $\hat{a}_{x}$ can be considered as a $c$-numbered quantity $\sqrt{n}e^{-i \varphi}$. Under this assumption with $\varphi=0$, we have $\hat{S}_{x}^{(I)} \sim n$, $\hat{S}_{y}^{(I)} \sim -\sqrt{n} (\hat{a}_{y}^{\dagger}+\hat{a}_{y})\equiv -\sqrt{2n}\hat{q}_{P}$, and $\hat{S}_{z}^{(I)} \sim -i \sqrt{n}(\hat{a}_{y}^{\dagger}-\hat{a}_{y})\equiv -\sqrt{2n}\hat{p}_{P}$. Thus the object's quantity, $\hat{L}_{y}$, is most significantly imprinted in 
\begin{equation}
\hat{S}_{z}^{(O)} \sim 2n \theta' \hat{L}_{y} + 2\sqrt{2n}\theta \hat{q}_{P}-\sqrt{2n}\hat{p}_{P}, \label{eq:homodyne0}
\end{equation}
where the last two terms are purely due to the probe photons and considered as the shot noise.

Putting the angular momentum operators also into the Jordan-Schwinger representation~\cite{UTK2006a,Leonhardt}, we have $\hat{L}_{x}=\frac{1}{2}(\hat{b}_{x}^{\dagger}\hat{b}_{x}-\hat{b}_{y}^{\dagger}\hat{b}_{y}) \sim N$, $\hat{L}_{y}=\frac{1}{2}(\hat{b}_{x}^{\dagger}\hat{b}_{y}+\hat{b}_{y}^{\dagger}\hat{b}_{x}) \sim \frac{\sqrt{N}}{2}(\hat{b}_{y}^{\dagger}+\hat{b}_{y})\equiv \sqrt{\frac{N}{2}}\hat{q}_{O}$, and $\hat{L}_{z}=\frac{1}{2i}(\hat{b}_{x}^{\dagger}\hat{b}_{y}-\hat{b}_{y}^{\dagger}\hat{b}_{x}) \sim i \frac{\sqrt{N}}{2}(\hat{b}_{y}^{\dagger}-\hat{b}_{y})\equiv \sqrt{\frac{N}{2}}\hat{p}_{O}$ because $\langle \hat{L}_{x} \rangle = N \gg1$ and thus the annihilation operator $\hat{b}_{x}$ can be considered as a $c$-numbered quantity $\sqrt{N}$. Here, the annihilation (creation) operators, $\hat{b}_{x}$ ($\hat{b}_{x}^{\dagger}$) and $\hat{b}_{y}$ ($\hat{b}_{y}^{\dagger}$), are the abstract ones, which subtract (add) $\hbar$ from (to) $\hat{L}_{x}$ and $\hat{L}_{y}$, respectively. $\hat{q}_{O}$ and $\hat{p}_{O}$ are the spinning objects' analogues of the quadrature operators. Measuring $\hat{S}_{z}^{(O)}$ in Eq.~(\ref{eq:homodyne0}) thus corresponds to measuring the quadrature of the object's angular momentum, i.e., 
\begin{equation}
\hat{S}_{z}^{(O)} \sim \sqrt{2}n\sqrt{N} \theta'\hat{q}_{O} + 2\sqrt{2n}\theta \hat{q}_{P}-\sqrt{2n}\hat{p}_{P}.
\label{eq:homodyne}
\end{equation}
This form is quite similar to that of the spin noise measurements~\cite{UTK2006a}, which indicates the possibility for measuring the angular momentum quantum noise of the birefringent dielectric object in a QND way~\cite{KBM1998epl,KMJYEB1999a,THTTIY1999a,UTK2006a} as well as for performing the vast related experiments, e.g., a generation of entangled spinning objects~\cite{JKP2001n}, a quantum memory for light with a spinning object~\cite{JSCFP2004n}, and quantum feedback control of object's angular momentum~\cite{GSM2004s}.

To get a ballpark figure of the signal-to-noise ratio (SNR) in measuring $\hat{q}_{O}$ for a micron-sized object, let us assign the presumed values in the case of the aforementioned quartz top to Eq.~(\ref{eq:homodyne}). For the visible light around $\lambda=600$~nm, the indices of refraction for quartz are $n_{e}=1.553$ and $n_{o}=1.544$~\cite{Fowles}, thus we have $\theta=\pi\frac{n_{e}-n_{o}}{\lambda}l \sim 0.094$. $\theta'=\frac{2\theta}{\langle \hat{L}_{x} \rangle}$ is then about $2.2\times10^{-12}$ with the top stabilized to rotate at 1~Hz, which has $\langle \hat{L}_{x} \rangle = N \sim 2.6\times10^{9}$ as mentioned before. The coefficient of the signal, $\hat{q}_{O}$, is then $5.1\times10^{-6} n $, whereas those of the noises, $\hat{q}_{P}$ and $\hat{p}_{P}$, are $0.27\sqrt{n}$ and $1.4\sqrt{n}$, respectively. Consequently, if the probe photon number $n$ can be made larger than $7.9\times10^{10}$~\cite{HAHLLMS2001op}, we can achieve the SNR greater than 1. In measuring the quadrature $\hat{q}_{O}$, the probe photons act as the local oscillator since the improvement of the SNR is proportional to $\sqrt{n}$. On the other hand, the SNR decreases at a rate proportional to $\sqrt{N}$, i.e., the square root of the angular momentum $\langle \hat{L}_{x} \rangle$. This is in a striking contrast with the spin quadrature measurement~\cite{UTK2006a}, in which the SNR increases with the square root of the spin counterpart of $\langle \hat{L}_{x} \rangle$~\cite{reason}. While for the SNR the lighter object is preferable to the heavier one because the coefficient for $\hat{q}_{O}$ in Eq.~(\ref{eq:homodyne}) grows with $\sqrt{N}\theta'=\frac{2\theta}{\sqrt{N}} \propto \frac{1}{\sqrt{\mu}}$, where $\mu$ is the total mass of the object, the object must be big enough so that the probe beam efficiently interacts with the object in the regime where the plane-wave approximation is valid.

Finally, let us discuss the practical issues for realizing the proposed experiment. First, it is crucial to suppress the classical angular momentum noise for measuring the intrinsic quantum noise. This requires a stable dipole force trap for the oblate symmetric top, a precise alignment of the extraordinary axis to the space-fixed x axis, and an appropriate stabilization of the angular momentum along the extraordinary axis. Second, the effect of the collision with the background gases should be taken into account to manage the decoherence of the rotational motion of the spinning object. Third, we should make the symmetric top as symmetrical as possible, otherwise the rotational motion may depart from the expected one. For the same reason, the extraordinary axis should be precisely matched with the principal axis of the object. We need, then, to consider to what extent we tolerate the deviation from the presumed perfect symmetric top. Systematic errors, e.g., the gravity and the inertial forces, must be considered, too. In reverse perspective the object would be used for the quantum-noise-limited inertial sensor.   

In conclusion, we show a scheme for observing the angular momentum quantum noise of a spinning micron-sized object in a QND way. The realization of the proposed scheme will, though challenging, provide an insight into the question of whether micron-sized objects exhibit quantum behaviors.

\begin{acknowledgments}
We thank Nobuyuki~Imoto, Yoshiaki~Kasahara, Yuki~Kawaguchi, Masahiro~Kitagawa, Mikio~Kozuma, Eugene~S.~Polzik, Takahiro~Sagawa, Akira~Shimizu, Shin~Takagi, and Masahito~Ueda for interesting discussions.
\end{acknowledgments}


\begin{thebibliography}{99}
\bibitem{reason}
The reason of this contrast can be given by the fact that the interaction term in Eq.~(\ref{eq:evolutionOp}) for measuring $\hat{L}_{y} \sim \sqrt{\frac{N}{2}}\hat{q}_{O}$, is, not the ideal spin-QND type, i.e., $\hat{S}_{y}\hat{L}_{y} \sim \sqrt{\frac{N}{2}}\hat{S}_{y}\hat{q}_{O}$, but $\hat{S}_{y}\hat{\phi}=\hat{S}_{y}\frac{\hat{L}_{y}}{\langle \hat{L}_{x} \rangle} \sim \sqrt{\frac{1}{2N}}\hat{S}_{y}\hat{q}_{O}$.
\end{thebibliography}
\end{document}